# Cognitive IoT based Health Monitoring Scheme using Non-Orthogonal Multiple Access

Ashiqur Rahman Rahul[1], Saifur Rahman Sabuj[1,2], Majumder Fazle Haider[1] and Shakil Ahmed[3]
[1]Department of Electrical and Electronic Engineering, BRAC University, Bangladesh
[2]Department of Electronics and Control Engineering, Hanbat National University, Korea
[3]Department of Electrical and Computer Engineering, The University of Arizona, AZ, USA
ashiqurrahul2506@gmail.com, s.r.sabuj@ieee.org, fazle.haider@bracu.ac.bd,
shakilahmed@ieee.org

**Abstract:** It has become very essential to address the limited spectrum capacity and their efficient utilization to support the increasing number of Internet of Things devices. When it comes to medical infrastructure, it becomes very imperative for medical devices to communicate with the base station. In such situations, communication over the wireless medium must provide optimized throughput (data rate) with effectual energy usage, which will ensure precise medical feedback by the responsible staff. Taking into account, it is necessary to operate wireless communication precisely at a higher frequency with more substantial bandwidth and low latency. Cognitive Radio (CR) is traditionally a viable choice, where it identifies and utilizes the vacant spectrum, thus maximizing the primary user's capacity and achieving spectral efficiency. To ensure such outcomes, the Non-Orthogonal Multiple Access (NOMA) techniques have proven to deliver an effective solution to the increasing number of devices with unimpaired performance, especially when the communication shifts towards a higher frequency band such as the mmWave band. In this chapter, IoT based CR network in uplink communication is proposed alongside employing NOMA techniques for optimal throughput, and energy efficiency for a medical infrastructure. Numerical results show that effectual throughput and energy efficiency for a High Reliable Communication (HRC) device and Moderate Reliable Communication (MRC) device improve over 83.13% and 73.95%, respectively and their corresponding energy efficacy values show vast improvement (83.11% and 73.96% respectively). Likewise, for interference case both the throughput and the energy efficiency improve approximately over 93% for all devices.

## 1. Introduction

Internet of Things (IoT), which can be implemented in every sphere of life, happens to be the most promising technology of fifth generation (5G) and beyond 5G wireless communications. Wearable sensors, when deployed for health care services to sense physical data of the human body and transmit the measured data to the nearest gateway wirelessly facilitating IoT, form a network called wireless body area network (WBAN). The concept of wireless technology integrated internet based health care services was first coined in [1]. The rapid advancement in the field of wireless technologies and wearable electronics in the recent time transform WBAN to a hot topic in the field of academia and industry to carry out top-notch research to bring out its full potential [2]. The main purpose of WBAN in health care services is the early detection of abnormalities through regular measuring physiological data. Thus, alerting the person to take necessary precautions and assist the medical professionals to make sound decisions regarding the appropriate treatment to cure diseases in the right time [3 – 5].

Due to the huge prospects of WBAN bringing revolutionary changes in telemedicine systems with real deployment of IoT, some different wireless standards have been proposed. IEEE wireless standard 802.15 defined as wireless personal area network (WPAN) is considered as suitable technology to fulfill the purpose of e-health systems [4]. IEEE 802.15 standard has several versions such as 802.15.1 and 802.15.4. These are commonly known as Bluetooth and ZigBee respectively, which are widely used as low power consumed, short ranged and low data transfer enabled wireless technology for infotainment and health care services [6]. Later on, IEEE originated 802.15.6 named as WBAN particularly developed for telemedicine systems with improvement in data reliability, sensor life time, latency and interference [7].

In the near future, a large number of sensors will be deployed on the body of the patients through WBANs to manage healthcare services remotely and provide intuitive decisions of medical professionals more dynamically. Therefore, handling big amounts of situational-awareness data in healthcare services for the future heterogeneous networks requires intelligent management of communication systems to ensure reliability, real time feedback and expected high quality services by utilizing minimum resources as much as possible. To meet up this challenge, cognitive IoT based wireless systems in WBANs for healthcare services can play a crucial role [8 - 10]. When a communication system, where the nearby transceivers cooperatively participate to ease the communication towards the destination and interact intelligently with humans to convey the instructions properly, is known as cognitive communication system. Some research works investigated the potential of IoT-cloud integrated communication framework for healthcare monitoring systems [11 – 13].

In this chapter, we have focused on designing cognitive IoT based wireless uplink communication systems by incorporating NOMA technique in WBANs for monitoring health conditions of categorized patients. The proposed system is intelligent enough to allocate high performance communication channels for exchanging information of the body nodes of Intensive Care Unit (ICU) patients, with high priority (HRC) and moderate performance communication (MRC) channels for regular patients based on computation of the Signal-to-Interference-plus-Noise (SINR) value. Theoretical analysis and simulation results of the performance metrics of such uplink communication validate the proposed optimized system to provide reliable data exchange applicable for healthcare services.

The remaining content of this chapter is structured accordingly: section 2 deals with the related research work in this sector. Next, section 3 describes the system model, which consists of network description, sensing and transmission process, pathloss model, evaluation of the mathematical model and optimization technique. Later section 4 illustrates the simulation outcomes through graphical representation, and finally, we conclude the chapter in section 5 by discussing the best energy efficiency case with some suggestions for future scopes of cognitive IoT.

## 2. Related Work

Several research works have been presented to highlight the prospects and challenges of WBAN and proposed optimal solutions to alleviate certain challenges. Inter-network interference is a common challenge that hinders the performance of WBANs when large numbers of wearable sensors act in a small dense area, a power control approach with low complexity based on game theory has been proposed in [14] to mitigate inter-network interference in WBANs. The authors focused on optimizing transmission power while keeping the system throughput high in mitigating inter-network interference in WBANs. The authors in [15] introduced the implications and advantages of human body communication PHY layer of IEEE 802.15.6 standard over other two PHY layer protocols due to its high conductivity of the human body on the performance of WBAN for health care services.

Few researchers have taken an attempt to implement the concept of energy harvesting in WBANs for telemedicine systems to prolong the lifetime of sensor nodes notably. In [16, 17], the authors analyzed power management strategy in energy harvested WBANs. A self-adaptive sensor has been proposed on TDMA frame structure to enable lifetime operation of the wearable sensors. The author in [18] presented a bandwidth allocation method for multiple WBANs along with beacon shifting and super frame interleaving integrated scheduling algorithm, which can ensure efficient bandwidth utilization in WBANs.

A good number of research works focusing on designing MAC layer protocols have been presented. In [19 - 21], the authors highlighted the significance of designing energy and delay aware MAC protocols to extend the durability of sensor nodes in WBAN. The authors in [22] developed SeDrip protocol incorporating Secure Hash Algorithm-1 based hash functions and Advanced Encryption Standard based encryption technique for highly secured and energy efficient data dissemination in WBANs. In [23, 24], the authors have demonstrated how the placement strategy of Body Node Coordinator (BNC) or hub in a WBAN can enhance the performance and lifetime of the body nodes notably. Moreover, the authors in [23] have shown that a suitable routing protocol plays a crucial role in prolonging the lifetime of body nodes and improves energy efficiency of a WBAN. The authors in [25, 26] introduced the impact of transmission delay of IEEE 802.15.6 CSMA/CA on duty-cycles of WBANs and computed the overall delay in WBANs using the theory of probability. A tele-medicine MAC layer protocol under the consideration of IEEE 802.15.4 standard CSMA/CA beacon mode enabled has been proposed in [27], the detailed analysis shows enhanced performance of the protocol in terms of delay, reliability and energy consumption. A hybrid MAC layer protocol has been proposed in [28] considering the advantages of both CSMA/CA and TDMA scheme simultaneously in order to enhance energy efficiency and also prolongs the lifetime of body nodes in WBANs. In [29 – 32], the authors have emphasized on different aspects of MAC layer protocols such as optimization of duty cycle and different coding techniques to analyze the energy and delay tradeoff in WBANs. Considering adaptive multi-dimensional traffic load and class, a MAC layer protocol has been proposed in [33], which can provide better performance in terms of energy efficiency and delay.

The authors in [34] discussed the potential security threats of WBANs through practical assessment and suggested implementing a forensic server and the use of hidden drones in the wireless network architecture to detect security threats. In [35], the authors demonstrated muscle strain sensor data measurement in a WBAN through prototype hardware development. The authors in [36] illustrated the deployment of magnetic induction based wireless communication systems for WBAN instead of typical electromagnetic wave communication technologies and showed impedance matching can significantly improve the efficiency of magnetic induction based WBAN.

The impacts of channel characterization on the performance of WBANs have been highlighted in few research works. In [37, 38], the authors addressed the necessity of Multi-Channel Broadcast (MCB) protocol to broadcast control signals from BNC to all the body nodes in WBANs. Additionally, by utilizing channel hopping sequences, the authors designed a MCB protocol, which can ensure minimum broadcast delay in asymmetric duty cycling and it is capable of broadcasting signal over multi-channels supported by IEEE 802.15.6 standard. The authors in [39] introduced channel allocation algorithm for typical WBANs using machine learning techniques. The proposed algorithm can allocate channels dynamically considering traffic load and provide optimum performance in terms of throughput, as well. In [40, 41], the authors analyzed the performance of WBANs by focusing on wireless channel modeling approaches. The authors in [40] highlighted the comparison of different types of relay systems performance on channel modeling and computed performance metrics to provide good insights. The authors in [41] investigated the performance of WBANs concentrating on the comparison of different channel modeling approaches for implanted and wearable sensors and calculated the bit error performance based on the movement of the human body.

On the other hand, the influence of gender and body shape on the performance of the body nodes in WBANs have been scrutinized in [42], the results showed to have higher pathloss and fading effect on males compared to females.

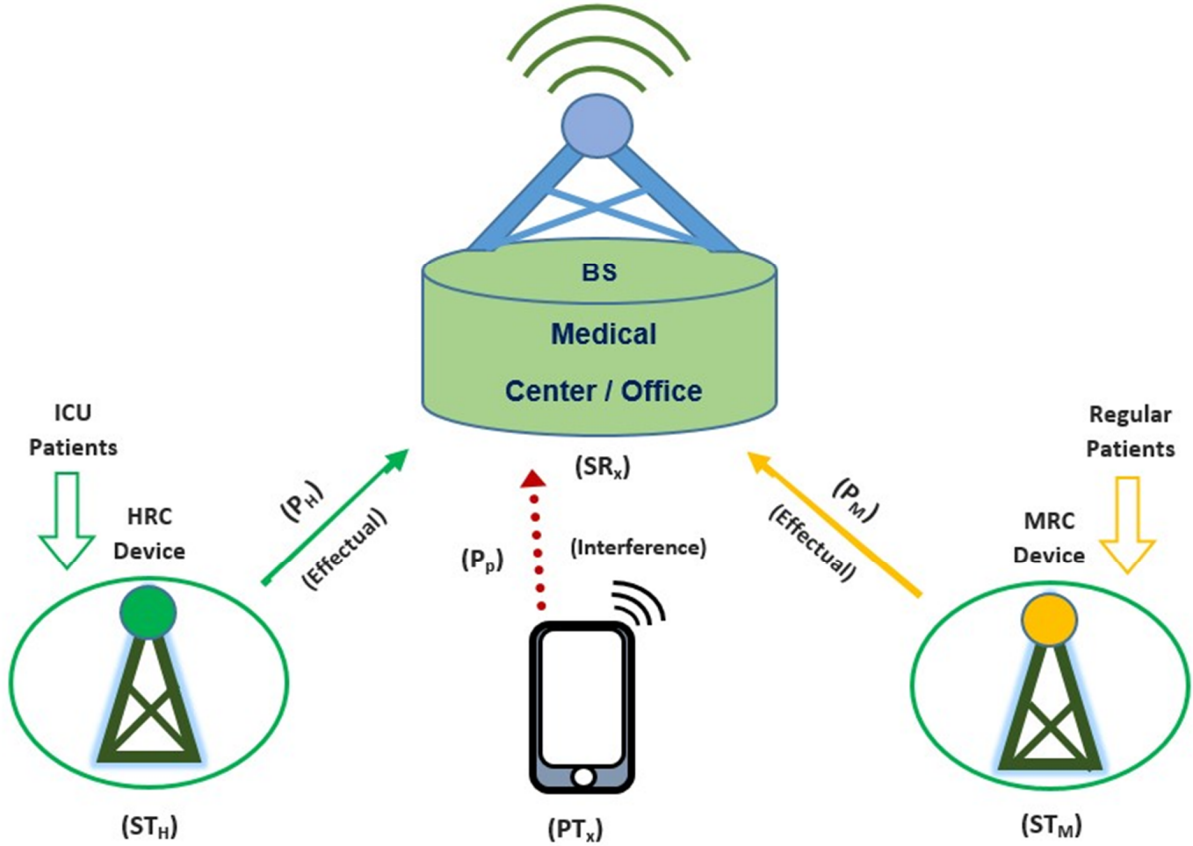

Figure 1. Communication Linkage of a Cognitive IoT based Health Monitoring Structure

## 3 System Model and Optimization

### 3.1 Network Description:

We have designed an uplink scenario for two types of devices concerning the health status of a patient with a cognitive IoT based communication network, shown in Fig. 1. Here, the distinguishing factor for the devices depends on the response of the medical staff toward a patient status. For instance, the HRC (High Reliable Communication) provides the medium of low latency with high-powered precise communication, for the patients needing immediate care. Conversely, MRC (Moderate Reliable Communication) devices have an average latency with low power, which deals with the conventional (non-urgent) responses towards regular patients. Moreover, in Fig. 1 the HRC secondary transmitter $(ST_H)$ and the MRC secondary transmitter $(ST_M)$ are in communication with the secondary receiver $(SR_x)$, which works as a base station (BS) in an uplink development. Additionally, $PT_x$ is the primary transmitter and it does not fall under the regulations of HRC or MRC devices, thus, in Fig. 1 $PT_x$ is an example of signal interference. Furthermore, in this figure, $P_H$, $P_M$, and $P_P$ are the powers necessary for transmission from $ST_H$, $ST_M$ and $PT_x$, respectively.

The HRC and MRC devices are utilizing NOMA methods within same Subcarrier ($S_C$) in an uplink communication [43]. The HRC and MRC devices use the same $S_C$ for communication, distinguished by their power levels. Thus, the $SR_x$ will perform Successive Interference Cancellation (SIC) to decode the particular signal apart from other signals in the same $S_C$.

### 3.2 Sensing and Transmission Process:

As sensing is an integral part of any cognitive device, our cognitive IoT system will also perform sensing operation on the radio spectrum to detect the underutilized channels. Both the secondary transmitters ($ST_H$ and $ST_M$) does the sensing process to identify any used or unused channels. This is established by comparing any received Signal-to-Noise Ratio (SNR) from any device to a threshold SNR. In our model if the received SNR ($\mu_P$) of $PT_x$ is lower than the threshold SNR ($\mu_{H/M}$) of HRC or MRC devices, then it is considered as an Effectual state. However, if the received SNR is higher, then $PT_x$ is considered to be in a transmitting state, which is denoted as the Interference state [44].

$$z = \begin{cases} 0, & when\ \mu_P < \mu_{H/M} \\ 1, & when\ \mu_P \geq \mu_{H/M} \end{cases} \quad (1)$$

Here, $z$ is an indicating factor for the active and inactive states of $PT_x$, where $z = 0$ indicates there is no interference ($\mu_P < \mu_{H/M}$) from the $PT_x$ meaning effectual state while HRC and MRC devices are transmitting and $z = 1$ indicates interference state ($\mu_P \geq \mu_{H/M}$) when $PT_x$ is active in transmission.

### 3.3 Pathloss Model:

Pathloss characterizes our system model as the attenuation of any signal strength, propagating through space between the transmitters ($ST_H, ST_M, PT_x$) and the receiver ($SR_x$). In this chapter, we have considered the channel gain for the interference state between the $PT_x$ and $SR_x$ as $g_p$. Likewise, for effectual states of HRC and MRC devices, channel gain between $ST_H$ and $SR_x$ is $g_h$, and between $ST_M$ and $SR_x$ channel gain is $g_m$. In terms of our model, we derived the channel gain characteristics for all communicative devices from the pathloss model established in [45, 46]. As proposed, our model considers the mmWave band for its uplink communication. Thus, it is possible to gain over 10-2000 m of distance from the secondary transmitters ($ST_H, ST_M$) to the BS, over 2-6 GHz of frequency, we can express the line-of-sight (LOS) pathloss ($Pl_{T2R}^{LOS}$), and non-line-of-sight (NLOS) pathloss ($Pl_{T2R}^{NLOS}$) as:

$$Pl_{T2R}^{LOS}(d_{T2R})[dB] = 22\ \log_{10}(d_{T2R}) + 28 + 20 \log_{10}(f_{c(mm)}) \quad (2)$$

$$Pl_{T2R}^{NLOS}(d_{T2R})[dB] = 36.7\ \log_{10}(d_{T2R}) + 22.7 + 26 \log_{10}(f_{c(mm)}) \quad (3)$$

Here, in the above expression T and R denotes all the transmitters and receiver in Fig. 1, where $d_{T2R}$ is their distance element. Additionally, $f_{c(mm)}$ exemplifies as the carrier frequency (2-6 GHz) over the mmWave band. Now, by utilizing the above expressions, we can develop the average pathloss over $d_{T2R}$ as:

$$Pl\ (d_{T2R}) = \omega * Pl_{T2R}^{LOS} + (1-\omega) * Pl_{T2R}^{NLOS} \qquad (4)$$

Where, $\omega$ is the coefficient probability of LOS link, over $d_{T2R}$ between all transmitters ($ST_H, ST_M, PT_x$) and BS.

### 3.4 Mathematical Model Evaluation:

This section will examine the performance metrics of HRC and MRC devices, such as their corresponding throughputs and energy efficiency. This will ultimately lead to the optimum solution for our system model through optimization techniques. Moreover, in our proposed model, we have assumed that, the HRC and MRC device observes the channel activity to sense the presence of $PT_x$. Since the presence of $PT_x$ results into two different types of throughputs at the receiver station $(SR_x)$, we establish the effectual and interference throughput (for both HRC and MRC device) with their corresponding energy efficiency aspect.

#### 3.4.1 Effectual Throughput:

As mentioned in the earlier section, if the presence of $PT_x$ is undiscovered, meaning only $ST_H$ and $ST_M$ is solely transmitting, then we denote it as an effectual state having effectual throughputs. Based on such states, we derive effectual throughputs for both HRC and MRC devices, by using $z = 0$ as an indicator for the absence of $PT_x$. Moreover, the effectual throughput will also imply as seamless spectrum sensing, since there is no interference from the $PT_x$. Hereafter, for $z = 0$, we can derive the effectual throughput $(S_H^0)$ for HRC device as:

$$S_H^0 = \left(\frac{t_t}{t_t + t_{se}}\right) p_x(z = 0)\ (1 - p_F) b \sum_{n=1}^{N_H} \log_2\left(1 + \frac{P_{H,n}\ |g_{h,n}|^2}{n_p\ b}\right) \qquad (5)$$

Where, $t_t$ and $t_{se}$ are transmission time and sensing time, respectively. $P_H$ is the power transmitted from $ST_H$ to $SR_x$ for any $n$ number of HRC device, $n_p$ is the noise power spectral density, $b$ is the bandwidth, and $g_h$ is the channel gain between $ST_H$ to $SR_x$ for any $n$ number of HRC device. Moreover, in the expression, $p_F$ is a probability output that exceeds a certain threshold when there is no signal (noise only), which represents an alarm of false detection. Next, $p_x(z = 0)$, is the probability of the inactive state of $PT_x$. Henceforth, we represent the term, $p_x(z = 0)\ (1 - p_F)$ as perfect detection probability, which implies that there is no interference from $PT_x$ between the channel $ST_H$ to $SR_x$.

Equally, for $z = 0$, we can derive the effectual throughput $(S_M^0)$ for any n number of MRC device as:

$$S_M^0 = \left(\frac{t_t}{t_t + t_{se}}\right) p_x(z = 0)\ (1 - p_F) b \sum_{n=1}^{N_M} \log_2\left(1 + \frac{P_{M,n}\ |g_{m,n}|^2}{n_p\ b + P_{H,n}\ |g_{h,n}|^2}\right) \qquad (6)$$

Here, $P_M$ is the transmission power from $ST_M$ to $SR_x$, and $g_m$ is the channel gain between them. Uniquely, for our uplink system model, higher power signal $(P_H |g_h|^2)$ in eq. (6) of any HRC device from $ST_H$ to $SR_x$, will pose diverse levels of interference to the lower power signals of any MRC device ($ST_M$ to $SR_x$).

### 3.4.2 Interference Throughput:

In a similar manner, we can derive throughputs for interference state. However, this time the HRC and MRC device senses the presence of $PT_x$, implying that $PT_x$ is also transmitting along with $ST_H$ and $ST_M$, causing an interference to the system. Hereafter, we can derive interference throughput $(S_H^1)$ for HRC device, using $z = 1$ as an indicator for the presence of $PT_x$ as:

$$S_H^1 = \left(\frac{t_t}{t_t + t_{se}}\right) p_x(z = 1)(1 - p_D) b \sum_{n=1}^{N_H} \log_2\left(1 + \frac{P_{H,n} |g_{h,n}|^2}{n_p b + P_P |g_P|^2}\right) \quad (7)$$

Here, $S_H^1$ is associated with imperfect spectrum sensing, since $PT_x$ is active on transmission. However, while comparing to non-HRC or non-MRC devices $S_H^1$ still provides the optimal throughput in presence of interference.

Moreover, in eq. (7), $P_P$ is the transmission power for $PT_x$, and $g_P$ is the channel gain between $PT_x$ and $SR_x$. Now, $p_D$ symbolizes the probability of sensing the presence of a targeted signal, meaning the probability output is less than a certain threshold, and $p_x(z = 1)$ represents the probability of $PT_x$ being active on transmission. Thus, $p_x(z = 1)(1 - p_D)$ is the imperfect detection probability for any HRC or MRC device, facing an interference from $PT_x$.

Similarly, for $z = 1$, we can derive the interference throughput $(S_M^1)$ for any n number of MRC device as:

$$S_M^1 = \left(\frac{t_t}{t_t + t_{se}}\right) p_x(z = 1)(1 - p_D) b \sum_{n=1}^{N_M} \log_2\left(1 + \frac{P_{M,n} |g_{m,n}|^2}{n_p b + P_{H,n} |g_{h,n}|^2 + P_P |g_P|^2}\right) \quad (8)$$

As for uplink communication, MRC devices will confront an added interference from the high-powered HRC device $(ST_H)$ and, the active primary transmitter $(PT_x)$. As per the decision made by IEEE 802.22 committee, when SNR $\leq -20$ dB, $p_D \geq 0.9$ and $p_F \leq 0.1$ [47]

### 3.4.3 Energy Efficiency:

In this chapter, we have defined the energy efficiency as the ratio of throughputs (i.e. effectual, and Interference) to the total power usage by the devices (i.e. HRC, and MRC). Hence, we can develop energy efficiency expression $(EE_H^z)$ for HRC device, in both effectual $(z = 0)$, and interference states $(z = 1)$ as:

$$EE_H^0 = \frac{S_H^0}{P_H + P_{cp} + P_{sp}} \quad (9)$$

$$EE_H^1 = \frac{S_H^1}{P_H + P_{cp} + P_{sp}} \quad (10)$$

Where, $P_H$ is the power required for transmission from $ST_H$ to $SR_x$, $P_{cp}$ is the power consumed by the HRC device circuit, and $P_{sp}$ is the power to sense the spectrum by the cognitive IoT based HRC device.

Similarly, we can establish, energy efficiency expression $(EE_M^z)$ for MRC device, in both effectual $(z = 0)$, and interference states $(z = 1)$ as:

$$EE_M^0 = \frac{S_M^0}{P_M + P_{cp} + P_{sp}} \quad (11)$$

$$EE_M^1 = \frac{S_M^1}{P_M + P_{cp} + P_{sp}} \quad (12)$$

Here, $P_M$ the transmission power for MRC device from $ST_M$ to $SR_x$, $P_{cp}$ and $P_{sp}$ is the circuit, spectrum sensing power for MRC devices.

### 3.4.4 Optimum Power :

In this subsection, we derive the expression for optimum power transmission in both effectual and interference state. Now, from the energy efficiency eq. for HRC and MRC device in eq. (9) and (11) at effectual state $(z = 0)$, we construct the following generalized single-objective optimization problems (SOP) as:

$$\max_{P_H} \ EE_H^0(P_H) \quad (13)$$

$$\max_{P_M} \ EE_M^0(P_M) \quad (14)$$

Likewise, for HRC and MRC device in interference state ($z = 1$), we can write with eq. (10) and (12) as:

$$\max_{P_H} \ EE_H^1(P_H) \qquad (15)$$

$$\max_{P_M} \ EE_M^1(P_M) \qquad (16)$$

Here in eq. (15) and (16), as interference from $PT_x$ is present, the energy efficiency value that we determine is the optimum energy efficiency.

### 3.4.4.1   Optimum Power derivation for HRC

Rewriting the eq. (9), we can obtain the optimum transmission power $\left(P_{H,n}^E\right)$ for the effectual state:

$$EE_H^0 = \frac{\left(\frac{t_t}{t_t + t_{se}}\right) p_x(z=0) (1 - p_F) b \sum_{n=1}^{N_H} \log_2\left(1 + \frac{P_{H.n}^E |g_{h,n}|^2}{n_p \, b}\right)}{P_H^E + P_{cp} + P_{sp}} \qquad (17)$$

Thus $P_{H.n}^E$ for any n number of HRC device is given by,

$$P_{H.n}^E = \frac{n_p \, b \left(\frac{P_{cp} \, g_{h.n}^2 + P_{sp} \, g_{h.n}^2 - n_p \, b}{W\left(0, \left(\frac{P_{cp} \, g_{h.n}^2 + P_{sp} \, g_{h.n}^2 - n_p \, b}{n_p \, b}\right) \exp(-1)\right)} - 1\right)}{g_{h.n}^2} \qquad (18)$$

*Proof:* Refer to Appendix 6.1

Where, $W(\cdot)$ is a Lambert function, which gives the solution to expression likely, $ae^a = b$. Here '$b$' acts as a function or a variable and '$a$' itself can be a constant or a complex variable, where '$a$' is acting both as an exponent and base. Moreover, we can solve such equations using $W$ to substitute any complex variable ($a$) by real variable ($b$). Henceforth, we can express the solution for such cases as: $W(b) = a$ [48].

Now, from eq. (10), we rewrite $EE_H^1$ for deriving optimum transmission power in interference state ($P_H^I$) as:

$$EE_H^1 = \frac{\left(\frac{t_t}{t_t + t_{se}}\right) p_x(z=1) (1 - p_D) b \sum_{n=1}^{N_H} \log_2\left(1 + \frac{P_{H.n}^I |g_{h,n}|^2}{n_p \, b + P_P \, |g_P|^2}\right)}{P_H^I + P_{cp} + P_{sp}} \qquad (19)$$

Thus resulting $P_{H.n}^I$ for any n number of HRC device is given by,

$$P_{H.n}^I = \frac{P_{cp}\, g_{h.n}^2 + P_{sp}\, g_{h.n}^2 - (n_p\, b + P_p\, g_p^2)}{W\left(0, \left(\frac{P_{cp}\, g_{h.n}^2 + P_{sp}\, g_{h.n}^2 - (n_p\, b + P_p\, g_p^2)}{n_p\, b + P_p\, g_p^2}\right) \exp(-1)\right) g_{h.n}^2} - \frac{(n_p\, b + P_p\, g_p^2)}{g_{h.n}^2} \quad (20)$$

*Proof:* Refer to Appendix 6.2

### 3.4.4.2 Optimum Power derivation for MRC

Correspondingly, rewriting the eq. (11), we can obtain the optimum transmission power $(P_{M,n}^E)$ for the effectual state:

$$EE_M^0 = \frac{\left(\frac{t_t}{t_t + t_{se}}\right) p_x(z=0)\, (1 - p_F)\, b \sum_{n=1}^{N_M} \log_2\left(1 + \frac{P_{M.n}^E\, |g_{m,n}|^2}{n_p\, b + P_{H,n}\, |g_{h,n}|^2}\right)}{P_M^E + P_{cp} + P_{sp}} \quad (21)$$

The resulting $P_{M.n}^E$ for any n number of MRC device is,

$$P_{M.n}^E = \frac{P_{cp}\, g_{m.n}^2 + P_{sp}\, g_{m.n}^2 - (n_p\, b + P_H\, g_{h.n}^2)}{W\left(0, \left(\frac{P_{cp}\, g_{m.n}^2 + P_{sp}\, g_{m.n}^2 - (n_p\, b + P_H\, g_{h.n}^2)}{n_p\, b + P_H\, g_{h.n}^2}\right) \exp(-1)\right) g_{m.n}^2} - \frac{(n_p\, b + P_H\, g_{h.n}^2)}{g_{m.n}^2} \quad (22)$$

*Proof:* Refer to Appendix 6.3

Next, we rewrite $EE_M^1$ of eq. (12) for deriving optimum transmission power in interference state $(P_{M,n}^I)$ as:

$$EE_M^1 = \frac{\left(\frac{t_t}{t_t + t_{se}}\right) p_x(z=1)\, (1 - p_D)\, b \sum_{n=1}^{N_M} \log_2\left(1 + \frac{P_{M.n}^I\, |g_{m,n}|^2}{n_p\, b + P_{H,n}\, |g_{h,n}|^2 + P_P\, |g_P|^2}\right)}{P_M^I + P_{cp} + P_{sp}} \quad (23)$$

Hence, the resulting $P_{M.n}^I$ for any n number of MRC device is,

$$P_{M.n}^I = \frac{P_{cp}\, g_{m.n}^2 + P_{sp}\, g_{m.n}^2 - (n_p\, b + P_H\, g_{h.n}^2 + P_p\, g_p^2)}{W\left(0, \left(\frac{P_{cp}\, g_{m.n}^2 + P_{sp}\, g_{m.n}^2 - (n_p\, b + P_H\, g_{h.n}^2 + P_p\, g_p^2)}{n_p\, b + P_H\, g_{h.n}^2 + P_p\, g_p^2}\right) exp(-1)\right) g_{m.n}^2}$$
$$- \frac{(n_p\, b + P_H\, g_{h.n}^2 + P_p\, g_p^2)}{g_{m.n}^2} \quad (24)$$

*Proof:* Refer to Appendix 6.4

As a result, to the optimum power expressions discussed above, throughput for HRC $(S_H^0)$ and MRC $(S_M^0)$ device is at an effectual state and has been improved and also the interference throughputs for HRC $(S_H^1)$ and MRC $(S_M^1)$ are improved.

## 4   *Simulation Results:*

To evaluate the performance of our system model, through Figures 2-5, we have applied eq. (18), (20), (22), and (24) to enhance both the throughput and energy efficiency of HRC and MRC devices in both effectual and interference state. There are five HRC and MRC devices in various distances. Next, the parameters selected for simulation are: $f_{c(mm)} = 5$ GHz, $b = 1$ MHz, $c = 3 \times 10^8$ m/sec, $t_t = 0.125 \times 10^{-3}$ sec, $t_{se} = 0.125 \times 10^{-3}$ sec, $p_x(z = 0) = 0:0.01:1$, $p_x(z = 1) = 0:0.01:1$, $p_F = 0.1$, $p_D = 0.9$, $n_p = -174$ dBm, $P_{cp} = 99$ dBm, $P_{sp} = 1$ dBm, $P_p = 50$ dBm, $P_H = (0.7 \times 30)$ dBm, and $P_M = (0.3 \times 30)$ dBm.

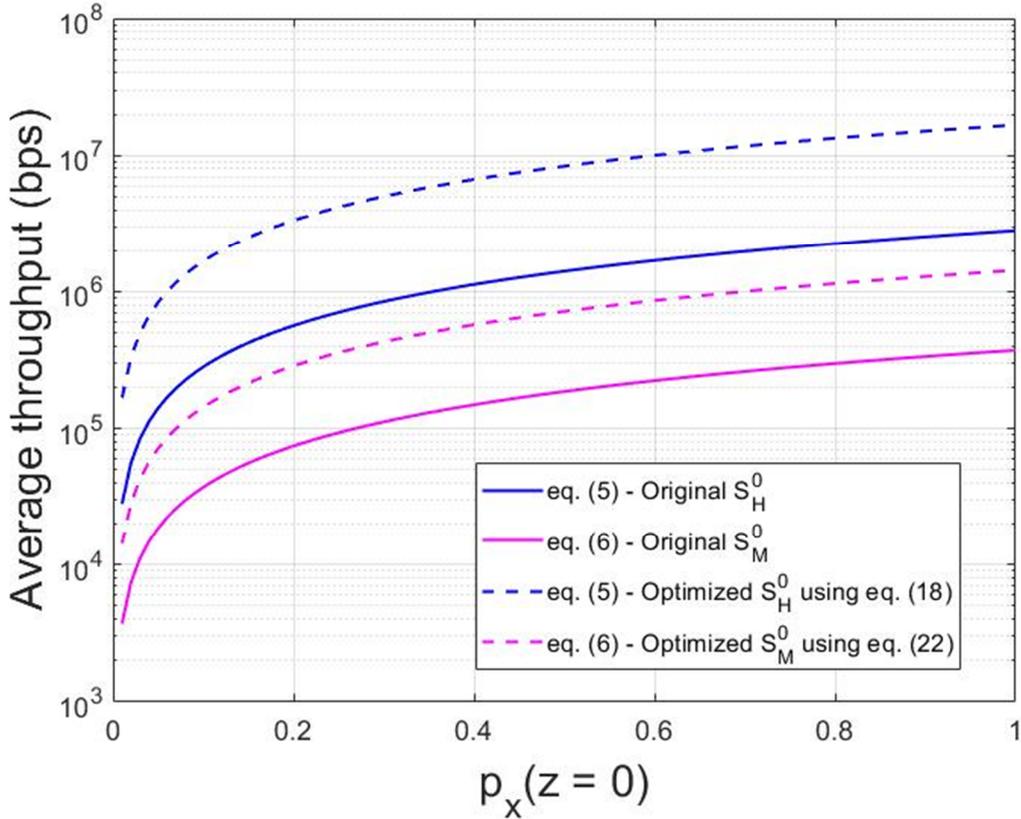

Figure 2. Average Effectual Throughput for both HRC and MRC device at numerous $p_x(z=0)$.

Figure 2 presents the average throughputs at effectual state $(z = 0)$ for both the HRC and MRC devices at several $p_x(z = 0)$. Now, $p_x(z = 0)$ represents that primary transmitter ($PT_x$) is not causing any interference to the system, and with more accurate $p_x(z = 0)$ (1 being the maximum value), throughputs for both HRC and MRC device will increase as well. Additionally, we enhance the original throughputs of HRC and MRC device in eq. (5) and (6) respectively by applying the optimum powers derived in eq. (18) and (19) for HRC and MRC devices. Thus, for a low latency and high-powered HRC device, original or conventional $S_H^0$ in eq. (5) at $p_x(z = 0) = 0.5$ is $1.409 \times 10^6$ bps and gradually increasing. Now, we enhance the throughput in eq. (5) by applying the optimum power $(P_H^E)$ from eq. (18) to acquire optimized $S_H^0$ for HRC device, which is $8.835 \times 10^6$ bps, at $p_x(z = 0) = 0.5$ and renders 83.13% optimized throughput over the original $S_H^0$. Similarly, for an MRC device requiring moderate power and latency, the original $S_M^0$ in eq. (6) is $1.864 \times 10^5$ bps and increasing, at $p_x(z = 0) = 0.5$, which is still considerably lower than the original $S_H^0$ in eq. (5). Later, we apply $P_M^E$ from eq. (22) to increase $S_M^0$ in eq. (6) such as, at $p_x(z = 0) = 0.5$ optimized throughput for MRC device is $7.157 \times 10^5$ bps, which is significantly better (73.96%) than the original $S_M^0$ in eq. (6). Furthermore, both the original and optimized throughputs for an MRC device $(S_M^0)$ will be lower than the average HRC throughputs $(S_H^0)$ since an HRC device requires higher power to produce greater throughput with low latency.

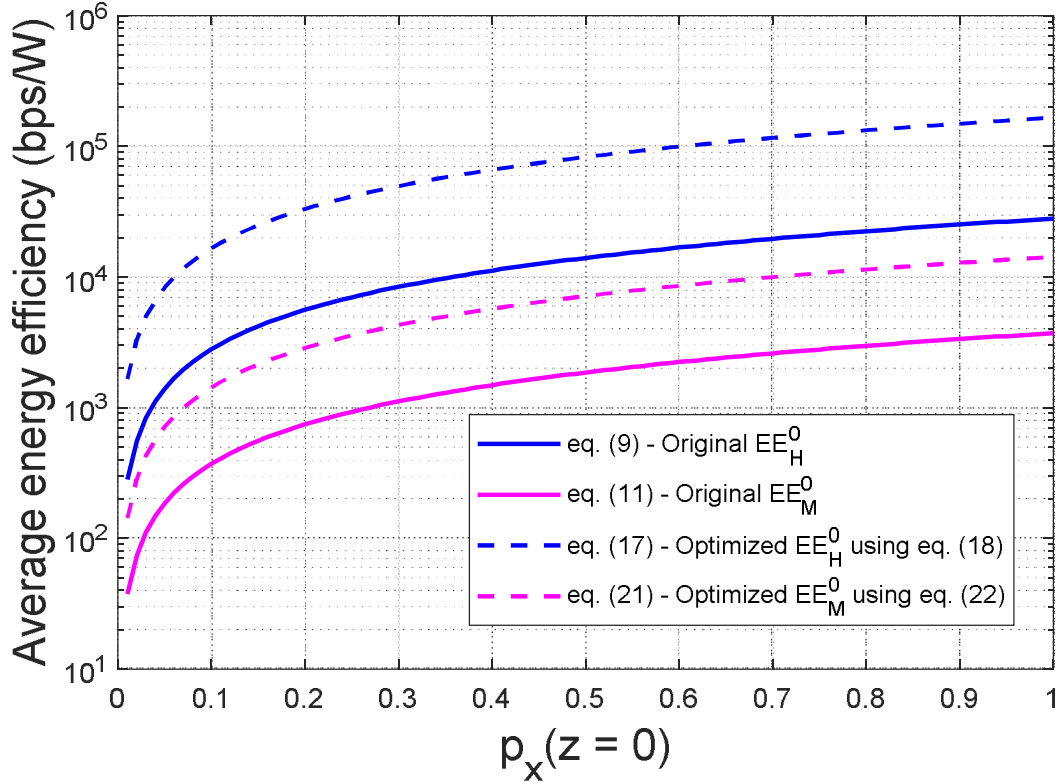

Figure 3. Average Effectual Energy Efficiency for both HRC and MRC device at various $p_x(z = 0)$.

Next, Figure 3 displays the effectual energy efficiency for both the HRC and MRC device at various $p_x(z = 0)$. As throughput for a device (HRC or MRC) increases with higher detection probability ($p_x(z = 0)$), the energy efficiency for that particular device will also rise if $p_x(z = 0)$ increases. Moreover, in Fig. 3, original energy efficiency ($EE_H^0$) from eq. (9) for an HRC device at effectual state is $1.4 \times 10^4$ bps/W at $p_x(z = 0) = 0.5$ and increasing progressively. We then improve this energy efficiency by applying $P_H^E$ from eq. (18) to obtain optimized energy efficiency ($EE_H^0$) in eq. (17), which is $8.3 \times 10^4$ bps/W at $p_x(z = 0) = 0.5$ and delivers 83.11% improvement over original $EE_H^0$. Likewise, in Fig. 3, original energy efficiency for an MRC device in eq. (11) gives 1858 bps/W at $p_x(z = 0) = 0.5$ and after applying $P_M^E$ from eq. (22), we obtain the improved energy efficiency ($EE_M^0$) in eq. (21). The optimized $EE_M^0$ is 7136 bps/W at $p_x(z = 0) = 0.5$, which gives 73% improvement over the original energy efficiency for an MRC device in eq. (11). Here also, both the original and optimized energy efficiency ($EE_H^0$) graphs for an HRC device is greater than all the energy efficiency levels of an MRC device since, throughput for HRC device is greater at all $p_x(z = 0)$ points.

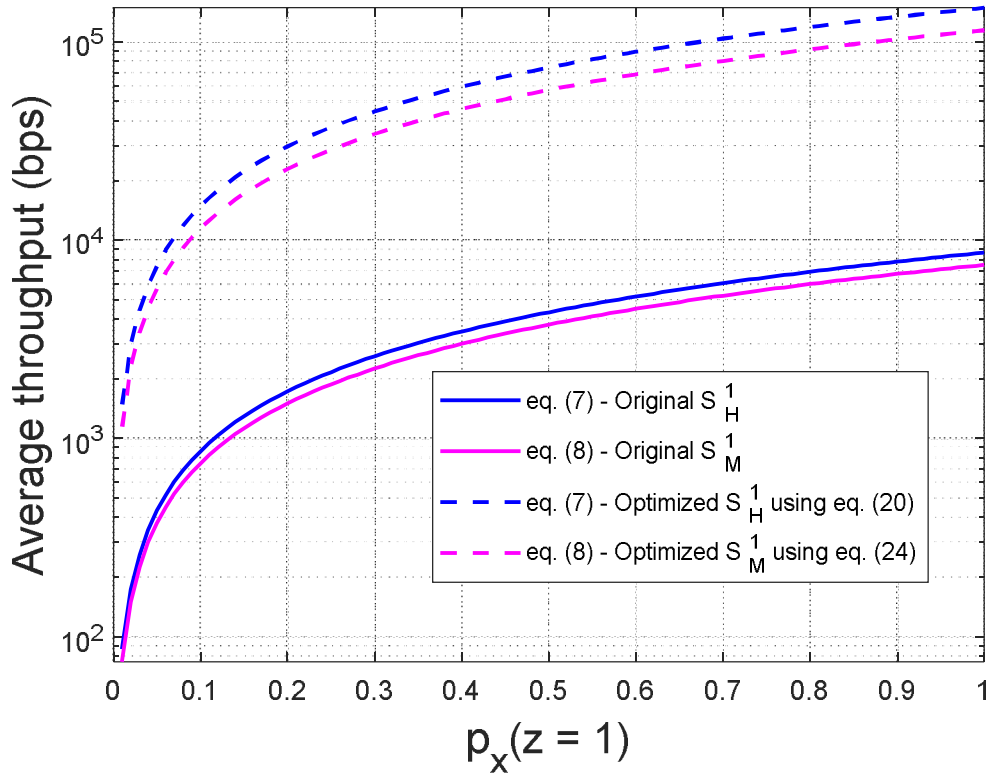

Figure 4. Average Interference Throughput for both HRC and MRC device at various $p_x(z = 1)$

Figure 4 now depicts the original and optimized throughputs of HRC and MRC devices when interference is present ($PT_x$ is active on transmission), with increasing $p_x(z = 1)$. Even in interference, higher detection probability $(p_x(z = 1))$ will provide greater throughputs. Now, for an HRC device facing interference, original throughput $(S_H^1)$ in eq. (7) is 4330 bps and original throughput $(S_M^1)$ for an MRC device in eq. (8) is 3753 bps, both at $p_x(z = 1) = 0.5$. Here, both the devices are moderately affected by interference resulting in more familiar throughputs but still distinguishable at the receiving end, as an HRC device will have higher throughputs with larger power for critical response towards a patient. Moreover, after applying optimum powers $(P_H^I$ and $P_M^I)$ from eq. (20) and (24) in eq. (7) and eq. (8), we acquire optimized throughputs for an HRC and MRC device, such as $7.419 \times 10^4$ bps and $5.72 \times 10^4$ bps, both at $p_x(z = 1) = 0.5$ respectivly. Further, in Fig. 4, it is clear that our optimized throughputs for both the HRC and MRC device are still superior when interference is present, and throughputs for the HRC and MRC device has improved by 94.16% and 93.43%, respectively over the original $S_H^1$ and $S_M^1$.

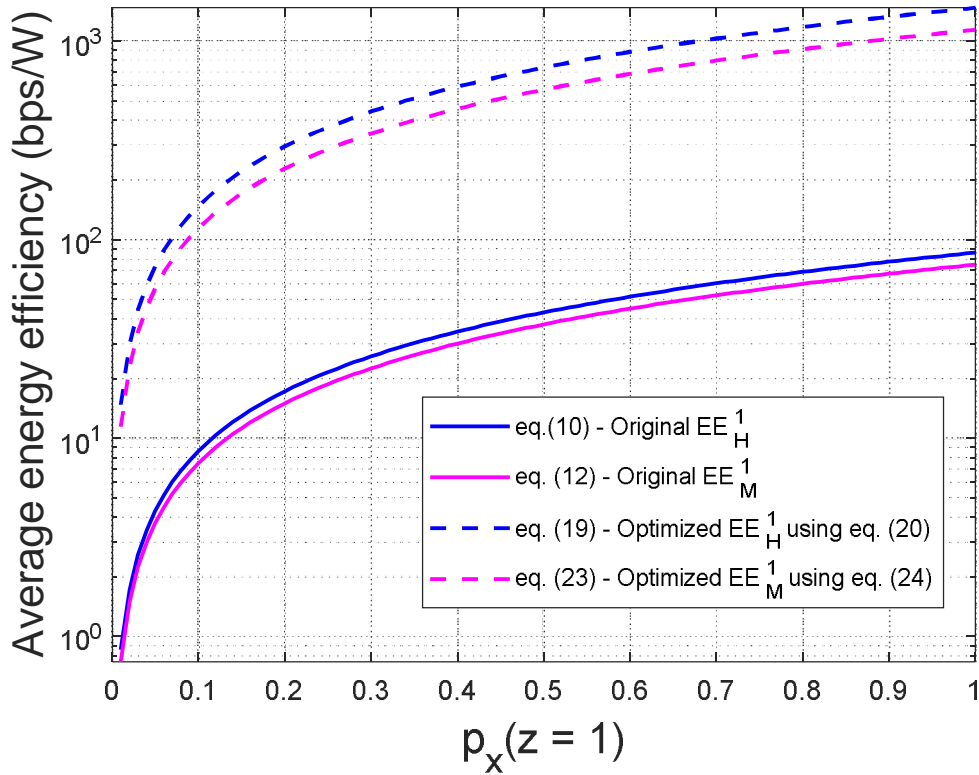

Figure 5. Average Interference Energy Efficiency for both HRC and MRC device at various $p_x(z = 1)$

Figure 5 finally displays the original and optimized energy efficiency for both the HRC and MRC devices at various $p_x(z = 1)$. Similarly, in Fig. 4, where interference throughputs for an HRC and MRC device are more familiar to each other, their resultant energy efficiencies (Fig. 5) will also be slightly familiar. Moreover, at $p_x(z = 1) = 0.5$, the original $EE_H^1$ in eq. (10) and $EE_M^1$ in eq. (12) are 42.99 bps/W and 37.41bps/W, respectively and they are increasing with detection that is more accurate ($p_x(z = 1) \rightarrow 1$). Later, we apply $P_H^I$ from eq. (20) and $P_M^I$ from eq. (24) to obtain optimized energy efficiencies for the HRC and MRC devices. Now from the Fig. 5, the optimized $EE_H^1$ in eq. (19) is 736.8 bps/W by 94.16% improvement over the original and optimized $EE_M^1$ is 570.3 bps/W by 93.44% improvement, for an HRC and MRC device respectively.

Overall, for both an HRC and MRC, the effectual throughputs and energy efficiency graphs are significantly greater than all the throughputs and energy efficiencies in the interference state. This is because at interference state, $PT_x$ is actively in communication with $SR_x$ causing external interference to the system. However, even in interference state, our optimized throughputs and energy efficiency still proves to be superior over the original. From Fig. 2-5, all the optimized results (Throughput and Energy efficiency) contributes more than 70% improvement over their original outcomes in uplink communication during both effectual and interference states. Henceforth, for a patient needing critical response or surgery, HRC devices will render efficient communication whereas, MRC devices will deliver adequate connection necessary for monitoring regular patients over both the effectual and interference state

## 5  *Conclusion:*

In conclusion, in this chapter, we introduced a cognitive IoT based system, which utilizes NOMA uplink communication to communicate successfully with the BS to execute, and process information depending on a patient's state of urgency. Here, we have generated the throughput expressions with their energy efficacy considering the presence and absent state of any interference. Later we have applied the derived optimum power expression to enhance both the throughput and energy efficiency for any HRC and MRC device. However, in the presence of any interference, our optimized throughput improves, by 94.16% and 93.43%, respectively for both HRC and MRC devices, with more than 93% improved energy efficiency collectively (both HRC and MRC). Similarly, for the effectual state, the throughput enhanced vastly, with 78.53% average (both HRC and MRC) improved energy efficiency. Hence, concerning the appropriate response toward patients, our system structure serves the optimum energy efficiency with or without any presence of interference. Moreover, concerning the massive usage of communicative IoT devices in all types of institutions, our system will provide a solid wireless communication medium over any cooperative network structure, which will ensure proper spectrum utilization [49]. Finally, our system already confirms significant strides in the energy efficiency factor, and we will further boost this with energy harvesting methods from natural resources that we will apply adjacent to wireless sensors for optimal energy-efficient performance [44, 50].

## 6  *Appendices:*

**6.1 Proof of Optimum Power Transmission for HRC Device at Effectual state $(z = 0)$:**

For the simplicity of the derivation, let us consider $P_{H.n} \rightarrow P_H$ & $g_{h,n} \rightarrow g_h$ for eq. (5). Now deriving $EE_H^0$ in eq. (9) with respect to $P_H$ we obtain,

$$\frac{dEE_H^0}{dP_H} = 0$$

$$\Rightarrow \frac{p_x(z=0)\, t_t\, (p_F - 1)\, b \log\left(1 + \frac{P_H\, g_h^2}{n_p\, b}\right)}{\log(2)\, (t_t + t_{se})\, (P_H + P_{cp} + P_{sp})^2}$$

$$- \frac{p_x(z=0)\, t_t\, (p_F - 1)\, g_h^2}{\log(2)\, (t_t + t_{se})\, n_p \left(1 + \frac{P_H\, g_h^2}{n_p\, b}\right)(P_H + P_{cp} + P_{sp})} = 0$$

$$\Rightarrow \frac{b \log\left(1 + \frac{P_H\, g_h^2}{n_p\, b}\right)}{P_H + P_{cp} + P_{sp}} = \frac{g_h^2}{n_p \left(1 + \frac{P_H\, g_h^2}{n_p\, b}\right)}$$

$$\Rightarrow \frac{\log\left(1 + \frac{P_H\, g_h^2}{n_p\, b}\right)}{P_H + P_{cp} + P_{sp}} = \frac{g_h^2}{b\left(n_p + \frac{P_H\, g_h^2}{b}\right)}$$

$$\Rightarrow \log\left(1 + \frac{P_H\, g_h^2}{n_p\, b}\right) = \frac{P_H\, g_h^2 + P_{cp}\, g_h^2 + P_{sp}\, g_h^2}{P_H\, g_h^2 + n_p\, b}$$

$$\Rightarrow \log\left(1 + \frac{P_H\, g_h^2}{n_p\, b}\right) = \frac{P_H\, g_h^2 + P_{cp}\, g_h^2 + P_{sp}\, g_h^2}{P_H\, g_h^2 + n_p\, b}$$

$$\Rightarrow \log\left(1 + \frac{P_H\, g_h^2}{n_p\, b}\right) - 1 = \frac{P_H\, g_h^2 + P_{cp}\, g_h^2 + P_{sp}\, g_h^2}{P_H\, g_h^2 + n_p\, b} - 1$$

$$\Rightarrow \log\left(1 + \frac{P_H\, g_h^2}{n_p\, b}\right) + \ln e^{-1} = \frac{P_{cp}\, g_h^2 + P_{sp}\, g_h^2 - n_p\, b}{P_H\, g_h^2 + n_p\, b}$$

$$\Rightarrow \log\left(\frac{P_H\, g_h^2 + n_p\, b}{n_p\, b}\right) \exp(-1) = \frac{P_{cp}\, g_h^2 + P_{sp}\, g_h^2 - n_p\, b}{P_H\, g_h^2 + n_p\, b}$$

$$\Rightarrow \left(\frac{P_H\, g_h^2 + n_p\, b}{n_p\, b}\right) \exp(-1) = \exp\left(\frac{P_{cp}\, g_h^2 + P_{sp}\, g_h^2 - n_p\, b}{P_H\, g_h^2 + n_p\, b}\right)$$

$$\Rightarrow \left(\frac{P_{cp}\, g_h^2 + P_{sp}\, g_h^2 - n_p\, b}{n_p\, b}\right) * \exp(-1)$$

$$= \left(\frac{P_{cp}\, g_h^2 + P_{sp}\, g_h^2 - n_p\, b}{P_H\, g_h^2 + n_p\, b}\right)$$

$$* \exp\left(\frac{P_{cp}\, g_h^2 + P_{sp}\, g_h^2 - n_p\, b}{P_H\, g_h^2 + n_p\, b}\right) \tag{6.25}$$

Applying the Lambert method to rewrite eq. (6.25) as: [a = be$^b$ $\Rightarrow$ W(a) = b]

$$\Rightarrow W\left(\left(\frac{P_{cp}\, g_h^2 + P_{sp}\, g_h^2 - n_p\, b}{n_p\, b}\right) \exp(-1)\right)$$

$$= \left(\frac{P_{cp}\, g_h^2 + P_{sp}\, g_h^2 - n_p\, b}{P_H\, g_h^2 + n_p\, b}\right)$$

$$\Rightarrow P_H \, g_h^2 + n_p \, b = \frac{P_{cp} \, g_h^2 + P_{sp} \, g_h^2 - n_p \, b}{W\left(\left(\frac{P_{cp} \, g_h^2 + P_{sp} \, g_h^2 - n_p \, b}{n_p \, b}\right) exp(-1)\right)}$$

$$\Rightarrow P_H = \frac{P_{cp} \, g_h^2 + P_{sp} \, g_h^2 - n_p \, b}{W\left(\left(\frac{P_{cp} \, g_h^2 + P_{sp} \, g_h^2 - n_p \, b}{n_p \, b}\right) exp(-1)\right) g_h^2} - \frac{n_p \, b}{g_h^2}$$

Thus, resulting $P_{H.n}^E$ for any 'n' number of HRC devices is –

$$\therefore P_{H.n}^E = \frac{n_p \, b \left(\frac{P_{cp} \, g_{h.n}^2 + P_{sp} \, g_{h.n}^2 - n_p \, b}{W\left(0, \left(\frac{P_{cp} \, g_{h.n}^2 + P_{sp} \, g_{h.n}^2 - n_p \, b}{n_p \, b}\right) exp(-1)\right)} - 1\right)}{g_{h.n}^2}$$

(6.26)

## 6.2 Proof of Optimum Power Transmission for HRC Device in Interference state $(z = 1)$:

For the simplicity of the derivation, let us consider $P_{H.n} \rightarrow P_H$ & $g_{h.n} \rightarrow g_h$ for eq. (7). Now deriving $EE_H^1$ in eq. (10) with respect to $P_H$ we attain,

$$\frac{dEE_H^1}{dP_H} = 0$$

$$\Rightarrow \frac{p_x(z=1) \, t_t \, (p_D - 1) \, b \log\left(1 + \frac{P_H \, g_h^2}{n_p \, b + P_p \, g_p^2}\right)}{\log(2) \, (t_t + t_{se}) \, (P_H + P_{cp} + P_{sp})^2}$$

$$- \frac{p_x(z=1) \, t_t \, (p_D - 1) \, b \, g_h^2}{\log(2) \, (t_t + t_{se}) \, (n_p \, b + P_p \, g_p^2) \left(1 + \frac{P_H \, g_h^2}{n_p \, b + P_p \, g_p^2}\right) (P_H + P_{cp} + P_{sp})}$$

$$= 0$$

$$\Rightarrow \frac{\log\left(1 + \frac{P_H \, g_h^2}{n_p \, b + P_p \, g_p^2}\right)}{P_H + P_{cp} + P_{sp}} = \frac{g_h^2}{(n_p \, b + P_p \, g_p^2)\left(1 + \frac{P_H \, g_h^2}{n_p \, b + P_p \, g_p^2}\right)}$$

$$\Rightarrow \log\left(1 + \frac{P_H \, g_h^2}{n_p \, b + P_p \, g_p^2}\right) - 1 = \frac{P_H \, g_h^2 + P_{cp} \, g_h^2 + P_{sp} \, g_h^2}{P_H \, g_h^2 + P_p \, g_p^2 + n_p \, b} - 1$$

$$\Rightarrow \log\left(1 + \frac{P_H\, g_h^2}{n_p\, b + P_p\, g_p^2}\right) + \ln e^{-1} = \frac{P_{cp}\, g_h^2 + P_{sp}\, g_h^2 - (n_p\, b + P_p\, g_p^2)}{P_H\, g_h^2 + P_p\, g_p^2 + n_p\, b}$$

$$\Rightarrow \log\left(\frac{P_H\, g_h^2 + P_p\, g_p^2 + n_p\, b}{n_p\, b + P_p\, g_p^2}\right)\exp(-1)$$
$$= \frac{P_{cp}\, g_h^2 + P_{sp}\, g_h^2 - (n_p\, b + P_p\, g_p^2)}{P_H\, g_h^2 + P_p\, g_p^2 + n_p\, b}$$

$$\Rightarrow \left(\frac{P_H\, g_h^2 + P_p\, g_p^2 + n_p\, b}{n_p\, b + P_p\, g_p^2}\right)\exp(-1)$$
$$= \exp\left(\frac{P_{cp}\, g_h^2 + P_{sp}\, g_h^2 - (n_p\, b + P_p\, g_p^2)}{P_H\, g_h^2 + P_p\, g_p^2 + n_p\, b}\right)$$

$$\Rightarrow \left(\frac{P_{cp}\, g_h^2 + P_{sp}\, g_h^2 - (n_p\, b + P_p\, g_p^2)}{n_p\, b + P_p\, g_p^2}\right) * \exp(-1)$$
$$= \left(\frac{P_{cp}\, g_h^2 + P_{sp}\, g_h^2 - (n_p\, b + P_p\, g_p^2)}{P_H\, g_h^2 + P_p\, g_p^2 + n_p\, b}\right)$$
$$* \exp\left(\frac{P_{cp}\, g_h^2 + P_{sp}\, g_h^2 - (n_p\, b + P_p\, g_p^2)}{P_H\, g_h^2 + P_p\, g_p^2 + n_p\, b}\right) \quad (6.27)$$

Utilizing the Lambert method to rewrite eq. (6.27) as: $[a = b e^b \Rightarrow W(a) = b]$

$$\therefore W\left(\left(\frac{P_{cp}\, g_h^2 + P_{sp}\, g_h^2 - (n_p\, b + P_p\, g_p^2)}{n_p\, b + P_p\, g_p^2}\right)\exp(-1)\right)$$
$$= \left(\frac{P_{cp}\, g_h^2 + P_{sp}\, g_h^2 - (n_p\, b + P_p\, g_p^2)}{P_H\, g_h^2 + P_p\, g_p^2 + n_p\, b}\right)$$

$$\Rightarrow P_H\, g_h^2 + P_p\, g_p^2 + n_p\, b$$
$$= \frac{P_{cp}\, g_h^2 + P_{sp}\, g_h^2 - (n_p\, b + P_p\, g_p^2)}{W\left(\left(\frac{P_{cp}\, g_h^2 + P_{sp}\, g_h^2 - (n_p\, b + P_p\, g_p^2)}{n_p\, b + P_p\, g_p^2}\right)\exp(-1)\right)}$$

Thus, resulting $P_{H.n}^I$ for any 'n' number of HRC devices is –

$$\therefore P_{H.n}^I = \frac{P_{cp}\, g_{h.n}^2 + P_{sp}\, g_{h.n}^2 - (n_p\, b + P_p\, g_p^2)}{W\left(0, \left(\frac{P_{cp}\, g_{h.n}^2 + P_{sp}\, g_{h.n}^2 - (n_p\, b + P_p\, g_p^2)}{n_p\, b + P_p\, g_p^2}\right)\exp(-1)\right) g_{h.n}^2}$$
$$- \frac{(n_p\, b + P_p\, g_p^2)}{g_{h.n}^2} \quad (6.28)$$

### 6.3 Proof of Optimum Power Transmission for MRC Device at Effectual state ($z = 0$):

For the simplicity of the derivation, let us consider $P_{M.n} \to P_M$ & $g_{m.n} \to g_m$ for eq. (6). Now deriving $EE_M^0$ in eq. (11) with respect to $P_M$ we obtain,

$$\frac{dEE_M^0}{dP_M} = 0$$

$$\Rightarrow \frac{p_x(z=0) \, t_t \, (p_F - 1) \, b \log\left(1 + \frac{P_M \, g_m^2}{n_p \, b + P_H \, g_h^2}\right)}{\log(2) \, (t_t + t_{se}) \, (P_M + P_{cp} + P_{sp})^2}$$

$$- \frac{p_x(z=0) \, t_t \, (p_F - 1) \, b \, g_m^2}{\log(2) \, (t_t + t_{se}) \, (n_p \, b + P_H \, g_h^2)\left(1 + \frac{P_M \, g_m^2}{n_p \, b + P_H \, g_h^2}\right)(P_M + P_{cp} + P_{sp})}$$

$$= 0$$

$$\Rightarrow \frac{\log\left(1 + \frac{P_M \, g_m^2}{n_p \, b + P_H \, g_h^2}\right)}{P_M + P_{cp} + P_{sp}} = \frac{g_m^2}{(n_p \, b + P_H \, g_h^2)\left(1 + \frac{P_M \, g_m^2}{n_p \, b + P_H \, g_h^2}\right)}$$

$$\Rightarrow \log\left(1 + \frac{P_M \, g_m^2}{n_p \, b + P_H \, g_h^2}\right) = \frac{P_M \, g_m^2 + P_{cp} \, g_m^2 + P_{sp} \, g_m^2}{P_M \, g_m^2 + P_H \, g_h^2 + n_p \, b}$$

$$\Rightarrow \log\left(1 + \frac{P_M \, g_m^2}{n_p \, b + P_H \, g_h^2}\right) - 1 = \frac{P_M \, g_m^2 + P_{cp} \, g_m^2 + P_{sp} \, g_m^2}{P_M \, g_m^2 + P_H \, g_h^2 + n_p \, b} - 1$$

$$\Rightarrow \log\left(1 + \frac{P_M \, g_m^2}{n_p \, b + P_H \, g_h^2}\right) + \ln e^{-1} = \frac{P_{cp} \, g_m^2 + P_{sp} \, g_m^2 - (n_p \, b + P_H \, g_h^2)}{P_M \, g_m^2 + P_H \, g_h^2 + n_p \, b}$$

$$\Rightarrow \left(\frac{P_M \, g_m^2 + P_H \, g_h^2 + n_p \, b}{n_p \, b + P_H \, g_h^2}\right) \exp(-1)$$

$$= \exp\left(\frac{P_{cp} \, g_m^2 + P_{sp} \, g_m^2 - (n_p \, b + P_H \, g_h^2)}{P_M \, g_m^2 + P_H \, g_h^2 + n_p \, b}\right)$$

$$\Rightarrow \left(\frac{P_{cp} \, g_m^2 + P_{sp} \, g_m^2 - (n_p \, b + P_H \, g_h^2)}{n_p \, b + P_H \, g_h^2}\right) * \exp(-1)$$

$$= \left(\frac{P_{cp} \, g_m^2 + P_{sp} \, g_m^2 - (n_p \, b + P_H \, g_h^2)}{P_M \, g_m^2 + P_H \, g_h^2 + n_p \, b}\right)$$

$$* \exp\left(\frac{P_{cp} \, g_m^2 + P_{sp} \, g_m^2 - (n_p \, b + P_H \, g_h^2)}{P_M \, g_m^2 + P_H \, g_h^2 + n_p \, b}\right) \tag{6.29}$$

Applying the Lambert method to rewrite eq. (6.29) as: $[a = be^b \Rightarrow W(a) = b]$

$$\therefore W\left(\left(\frac{P_{cp} \, g_m^2 + P_{sp} \, g_m^2 - (n_p \, b + P_H \, g_h^2)}{n_p \, b + P_H \, g_h^2}\right) \exp(-1)\right)$$

$$= \left(\frac{P_{cp} \, g_m^2 + P_{sp} \, g_m^2 - (n_p \, b + P_H \, g_h^2)}{P_M \, g_m^2 + P_H \, g_h^2 + n_p \, b}\right)$$

$$\Rightarrow P_M\, g_m^2 + P_H\, g_h^2 + n_p\, b$$
$$= \frac{P_{cp}\, g_m^2 + P_{sp}\, g_m^2 - (n_p\, b + P_H\, g_h^2)}{W\left(\left(\dfrac{P_{cp}\, g_m^2 + P_{sp}\, g_m^2 - (n_p\, b + P_H\, g_h^2)}{n_p\, b + P_H\, g_h^2}\right)\exp(-1)\right)}$$

Thus, resulting $P_{M.n}^E$ for any 'n' number of MRC devices is –

$$\therefore P_{M.n}^E = \frac{P_{cp}\, g_{m.n}^2 + P_{sp}\, g_{m.n}^2 - (n_p\, b + P_H\, g_{h.n}^2)}{W\left(0,\left(\dfrac{P_{cp}\, g_{m.n}^2 + P_{sp}\, g_{m.n}^2 - (n_p\, b + P_H\, g_{h.n}^2)}{n_p\, b + P_H\, g_{h.n}^2}\right)\exp(-1)\right) g_{m.n}^2} - \frac{(n_p\, b + P_H\, g_{h.n}^2)}{g_{m.n}^2}$$

(6.30)

## 6.4 Proof of Optimum Power Transmission for MRC Device in Interference state $(z = 1)$:

For the simplicity of the derivation, let us consider $P_{M.n} \to P_M$ & $g_{m.n} \to g_m$ for eq. (8). Now deriving $EE_M^0$ in eq. (12) with respect to $P_M$ we obtain,

$$\frac{dEE_M^1}{dP_M}$$

$$\Rightarrow \frac{p_x(z=1)\, t_t\, (p_D - 1)\, b \log\left(1 + \dfrac{P_M\, g_m^2}{n_p\, b + P_H\, g_h^2 + P_p\, g_p^2}\right)}{\log(2)\, (t_t + t_{se})\, (P_M + P_{cp} + P_{sp})^2}$$

$$- \frac{p_x(z=1)\, t_t\, (p_D - 1)\, b\, g_m^2}{\log(2)\, (t_t + t_{se})\, (n_p\, b + P_H\, g_h^2 + P_p\, g_p^2)\left(1 + \dfrac{P_M\, g_m^2}{n_p\, b + P_H\, g_h^2 + P_p\, g_p^2}\right) (P_M + P_{cp} + P_{sp})}$$

$$= 0$$

$$\Rightarrow \frac{\log\left(1 + \dfrac{P_M\, g_m^2}{n_p\, b + P_H\, g_h^2 + P_p\, g_p^2}\right)}{P_M + P_{cp} + P_{sp}}$$

$$= \frac{g_m^2}{(n_p\, b + P_H\, g_h^2 + P_p\, g_p^2)\left(1 + \dfrac{P_M\, g_m^2}{n_p\, b + P_H\, g_h^2 + P_p\, g_p^2}\right)}$$

$$\Rightarrow \log\left(1 + \frac{P_M\, g_m^2}{n_p\, b + P_H\, g_h^2 + P_p\, g_p^2}\right) - 1 = \frac{P_M\, g_m^2 + P_{cp}\, g_m^2 + P_{sp}\, g_m^2}{P_M\, g_m^2 + P_H\, g_h^2 + P_p\, g_p^2 + n_p\, b} - 1$$

$$\Rightarrow \log\left(1 + \frac{P_M\, g_m^2}{n_p\, b + P_H\, g_h^2 + P_p\, g_p^2}\right) + \ln e^{-1}$$

$$= \frac{P_{cp}\, g_m^2 + P_{sp}\, g_m^2 - (n_p\, b + P_H\, g_h^2 + P_p\, g_p^2)}{P_M\, g_m^2 + P_H\, g_h^2 + P_p\, g_p^2 + n_p\, b}$$

$$\Rightarrow \left(\frac{P_M\, g_m^2 + P_H\, g_h^2 + P_p\, g_p^2 + n_p\, b}{n_p\, b + P_H\, g_h^2 + P_p\, g_p^2}\right) \exp(-1)$$

$$= \exp\left(\frac{P_{cp}\, g_m^2 + P_{sp}\, g_m^2 - (n_p\, b + P_H\, g_h^2 + P_p\, g_p^2)}{P_M\, g_m^2 + P_H\, g_h^2 + P_p\, g_p^2 + n_p\, b}\right)$$

$$\Rightarrow \left(\frac{P_{cp}\, g_m^2 + P_{sp}\, g_m^2 - (n_p\, b + P_H\, g_h^2 + P_p\, g_p^2)}{n_p\, b + P_H\, g_h^2 + P_p\, g_p^2}\right) * \exp(-1)$$

$$= \left(\frac{P_{cp}\, g_m^2 + P_{sp}\, g_m^2 - (n_p\, b + P_H\, g_h^2 + P_p\, g_p^2)}{P_M\, g_m^2 + P_H\, g_h^2 + P_p\, g_p^2 + n_p\, b}\right)$$

$$* \exp\left(\frac{P_{cp}\, g_m^2 + P_{sp}\, g_m^2 - (n_p\, b + P_H\, g_h^2 + P_p\, g_p^2)}{P_M\, g_m^2 + P_H\, g_h^2 + P_p\, g_p^2 + n_p\, b}\right) \quad (6.31)$$

Utilizing the Lambert method to rewrite eq. (6.31) as: [a = be$^b$ $\Rightarrow$ W(a) = b]

$$\Rightarrow W\left(\left(\frac{P_{cp}\, g_m^2 + P_{sp}\, g_m^2 - (n_p\, b + P_H\, g_h^2 + P_p\, g_p^2)}{n_p\, b + P_H\, g_h^2 + P_p\, g_p^2}\right) \exp(-1)\right)$$

$$= \left(\frac{P_{cp}\, g_m^2 + P_{sp}\, g_m^2 - (n_p\, b + P_H\, g_h^2 + P_p\, g_p^2)}{P_M\, g_m^2 + P_H\, g_h^2 + P_p\, g_p^2 + n_p\, b}\right)$$

$$\Rightarrow P_M\, g_m^2 + P_H\, g_h^2 + P_p\, g_p^2 + n_p\, b$$

$$= \frac{P_{cp}\, g_m^2 + P_{sp}\, g_m^2 - (n_p\, b + P_H\, g_h^2 + P_p\, g_p^2)}{W\left(\left(\frac{P_{cp}\, g_m^2 + P_{sp}\, g_m^2 - (n_p\, b + P_H\, g_h^2 + P_p\, g_p^2)}{n_p\, b + P_H\, g_h^2 + P_p\, g_p^2}\right) \exp(-1)\right)}$$

Finally, the resulting $P_{M.n}^I$ for any 'n' number of MRC devices is –

$$P_{M.n}^I = \frac{P_{cp}\, g_{m.n}^2 + P_{sp}\, g_{m.n}^2 - (n_p\, b + P_H\, g_{h.n}^2 + P_p\, g_p^2)}{W\left(0, \left(\frac{P_{cp}\, g_{m.n}^2 + P_{sp}\, g_{m.n}^2 - (n_p\, b + P_H\, g_{h.n}^2 + P_p\, g_p^2)}{n_p\, b + P_H\, g_{h.n}^2 + P_p\, g_p^2}\right) \exp(-1)\right) g_{m.n}^2}$$

$$- \frac{(n_p\, b + P_H\, g_{h.n}^2 + P_p\, g_p^2)}{g_{m.n}^2} \quad (6.32)$$

Systems," IEEE Access, vol. 4, pp. 6325 - 6343, 2016.